\documentclass[sigconf, hyphens, obeyspaces]{acmart}

\usepackage{listings}
\usepackage[utf8]{inputenc}
\usepackage{rotating}
\usepackage[acronym]{glossaries}
\usepackage{booktabs}
\usepackage{tikz}
\usepackage{wasysym}
\usepackage{pifont}
\usepackage[noabbrev, capitalize]{cleveref}
\usepackage{makecell}

\setacronymstyle{long-short}

\def\BibTeX{{\rm B\kern-.05em{\sc i\kern-.025em b}\kern-.08emT\kern-.1667em\lower.7ex\hbox{E}\kern-.125emX}}

\title{Grand Theft App: Digital Forensics of Vehicle Assistant Apps}
\setcopyright{none}
\renewcommand\footnotetextcopyrightpermission[1]{}

\author{Simon Ebbers}
\affiliation{%
  \institution{Münster University of Applied Sciences}
  \streetaddress{Stegerwaldstr. 39} 
  \city{Steinfurt} 
  \country{Germany}
}

\author{Fabian Ising}
\affiliation{%
  \institution{Münster University of Applied Sciences}
  \streetaddress{Stegerwaldstr. 39} \city{Steinfurt} \country{Germany}
}

\author{Christoph Saatjohann}
\affiliation{%
  \institution{Münster University of Applied Sciences}
  \streetaddress{Stegerwaldstr. 39} \city{Steinfurt} \country{Germany}
}

\author{Sebastian Schinzel}
\affiliation{%
  \institution{Münster University of Applied Sciences}
  \streetaddress{Stegerwaldstr. 39} \city{Steinfurt} \country{Germany}
}

\newcommand{\unknown}{\textcolor{gray}{$\varnothing$}}
\newcommand{\cached}{\LEFTcircle}
\newcommand{\partialdata}{\LEFTcircle}
\newcommand{\nodata}{\hspace{0.1em}--}
\newcommand{\encrypted}{\Circle}
\newcommand{\metadata}{\Circle}
\newcommand{\available}{\CIRCLE}
\newcommand{\additional}{\ding{58}}

\usepackage{xparse}

\ExplSyntaxOn
\NewDocumentCommand{\replace}{mmm}
 {
  \marian_replace:nnn {#1} {#2} {#3}
 }

\tl_new:N \l_marian_input_text_tl

\cs_new_protected:Npn \marian_replace:nnn #1 #2 #3
 {
  \tl_set:Nn \l_marian_input_text_tl { #1 }
  \tl_replace_all:Nnn \l_marian_input_text_tl { #2 } { #3 }
  \tl_use:N \l_marian_input_text_tl
 }
\ExplSyntaxOff
\let\OldTexttt\texttt
\renewcommand{\texttt}[1]{\OldTexttt{\replace{#1}{.}{.\allowbreak}}}

\renewcommand\theadfont{\small\bfseries}

\begin{document}

\begin{abstract}
Due to the increasing connectivity of modern vehicles, collected data is no longer only stored in the vehicle itself but also transmitted to car manufacturers and vehicle assistant apps. This development opens up new possibilities for digital forensics in criminal investigations involving modern vehicles. This paper deals with the digital forensic analysis of vehicle assistant apps of eight car manufacturers.
We reconstruct the driver's activities based on the data stored on the smartphones and in the manufacturer's backend.  

For this purpose, data of the Android and iOS apps of the car manufacturers Audi, BMW, Ford, Mercedes, Opel, Seat, Tesla, and Volkswagen were extracted from the smartphone and examined using digital forensic methods in accordance with lawful government-approved forensics guidelines. Additionally, manufacturer data was retrieved using Subject Access Requests.
Using the extensive data gathered, we successfully reconstruct trips and refueling processes, determine parking positions and duration, and track the locking and unlocking of the vehicle.

These findings show that the digital forensic investigation of smartphone applications is a useful addition to vehicle forensics and should therefore be taken into account in the strategic preparation of future digital forensic investigations.
\end{abstract}

\twocolumn
\maketitle
\pagestyle{empty}

\newacronym{BSI}{BSI}{German Federal Office for Information Security}
\newacronym{NIST}{NIST}{National Institute of Standards and Technology}
\newacronym{CFSAP}{CFSAP}{Computer Forensic - Secure, Analyze, Present}
\newacronym{plist}{plist}{Property List}
\newacronym{ffs}{FFS}{Full File System}
\newacronym{twrp}{TWRP}{Team Win Recovery Project}
\newacronym{vin}{VIN}{vehicle identification number}
\newacronym{obd}{OBD}{On-Board Diagnostic}
\newacronym{GDPR}{GDPR}{General Data Protection Regulation}
\newacronym{gsm}{GSM}{Global System for Mobile Communications}
\newacronym{uuid}{UUID}{Universally Unique Identifier}
\newacronym{JSON}{JSON}{JavaScript Object Notation}
\newacronym{HTML}{HTML}{Hypertext Markup Language}
\newacronym{JPEG}{JPEG}{Joint Photographics Expert Group}
\newacronym{blob}{BLOB}{Binary Large Object}
\newacronym{url}{URL}{Uniform Resource Locator}
\newacronym{xml}{XML}{Extensible Markup Language}
\newacronym{imei}{IMEI}{International Mobile Station Equipment Identity}
\newacronym{iccid}{ICCID}{Integrated Circuit Card Identifier}
\newacronym{imsi}{IMSI}{International Mobile Subscriber Identity}
\newacronym{msisdin}{MSISDIN}{Mobile Station Integrated Services Digital Network Number}
\newacronym{ncap}{Euro NCAP}{European New Car Assessment Programme}
\newacronym{sar}{SAR}{Subject Access Request}

\section{Introduction}
This is the extended version of the Short Paper \textit{Grand Theft App: Digital Forensics of Vehicle Assistant Apps} published at the 16th International Conference on Availability, Reliability and Security (ARES 2021).
\vspace{\baselineskip}

\begin{figure}[t]
    \centering
    \includegraphics[width=0.7\columnwidth]{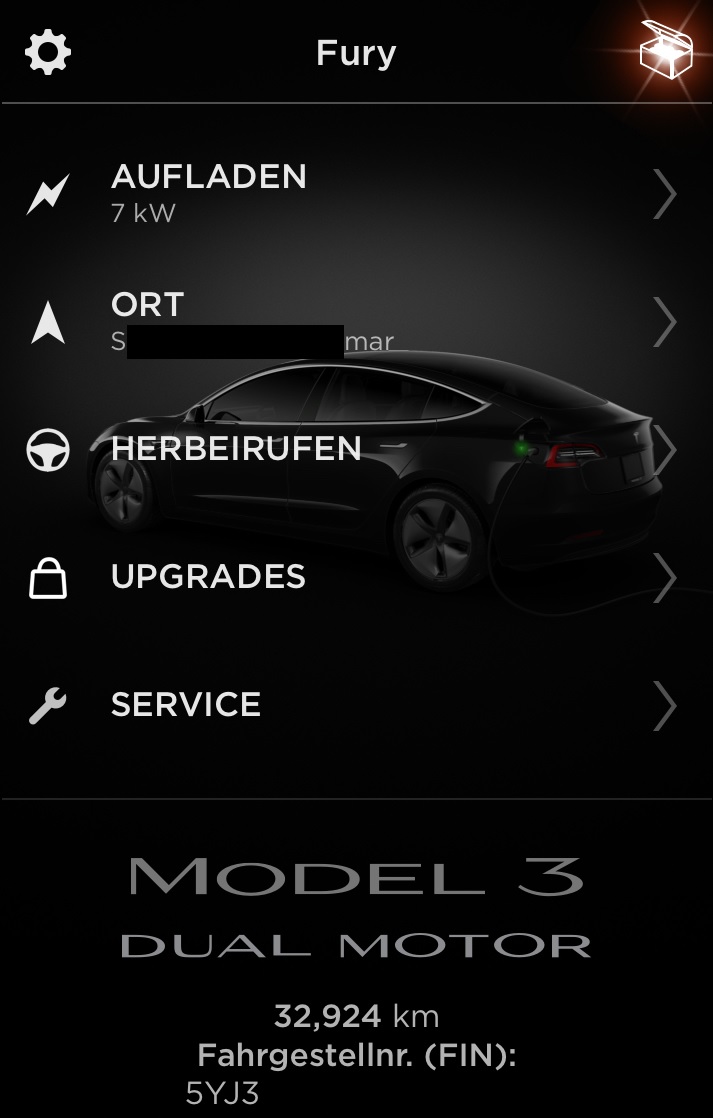}
    \caption{Screenshot of the Tesla iOS app. The current location, as well as the VIN (both redacted), are displayed.}
    \label{fig:screenshot}
\end{figure}

Thanks to the large number of assistance systems and the possibilities offered by infotainment systems, modern vehicles give the impression of being computers on wheels. In the past, smart connected systems were reserved for luxury-class vehicles, but nowadays, they are introduced even into compact class vehicles. %
On the other hand, the luxury-class is heading towards automated driving \cite{9046805}, introducing a wide range of sensors to vehicles. 
Interfaces between the vehicle network and the infotainment system allow sensor data to be checked not only from inside the vehicle via the on-board computer but also via the corresponding app on the smartphone \cite{adasstatistic}. It becomes apparent from the screenshot of the Tesla iOS app displayed in \cref{fig:screenshot} that these apps handle potentially sensitive data -- e.g., location data.

This development is also being driven forward by \gls{ncap}, which evaluates the safety of newly launched cars. It requires state-of-the-art adaptive cruise control and lane departure warning systems with additional functions supporting the driver and keeping them alert to reach the highest safety rating \cite{ncap_explanation}. In addition to these evaluation criteria for safety, the EU-mandated automatic emergency call system eCall requires car manufacturers to install telecommunications technology in vehicles \cite{ecall}.

This development and the associated collection and storage of sensitive data pose new opportunities for criminal investigations. For example, in 2017, investigators used time-stamped voice commands from a car infotainment system to prove a murder case \cite{insecurewheels}. While vehicle data is slowly becoming an essential source of information for law enforcement, investigations of the data generated from vehicle assistant apps are far less common. To evaluate the exchange and storage of this sensitive data, it is necessary to clarify which and where data is stored and which options exist for criminal investigators to access it.

In this research, we consider a twofold approach to this data acquisition. Firstly, we assume that car manufacturers store their users' data and hand them over to investigating authorities where required by law. In the European Union, this obligation is covered by the \gls{GDPR} (Article 23) \cite{gdpr}. Secondly, the fact that smartphone apps can access the vehicles suggests that data about the respective vehicles are also stored on the connected smartphones. Therefore, we aim to answer the following research question: Which data can be forensically acquired from vehicle assistant applications or requested from the manufacturer, and which activities of the driver can be reconstructed based on this data?

To answer this question, we generate data using a specific usage profile, retrieve the data from smartphone storage using digital forensic methods, and request data from the manufacturer using \glspl{sar}. We then analyze the acquired data to reconstruct the driver's actions.

We analyzed ten popular vehicle assistant apps on both iOS and Android, focusing on six information categories. We found that all tested apps leave forensic data traces relevant to the driver's behavior on the phone storage.

\paragraph{Contributions}

We make the following contributions:
\begin{itemize}
    \item We introduce a new data source for criminal investigations by combining forensic smartphone and vehicle data generated by vehicle assistant apps.
    \item We provide the first public structured analysis of the forensic data left behind on smartphone storage by vehicle assistant apps.
    \item We analyze which data is stored by manufacturers of connected vehicles using Subject Access Requests and its relation to the driver's actions.
    \item We analyze the combined data of ten popular vehicle assistant apps for forensic data that potentially supports criminal investigations.
\end{itemize}

\section{Related Work and Background}
\subsection{Digital forensic background}

The locard's rule, taken from criminology, states that no perpetrator can commit a crime or leave a crime scene without leaving a multitude of traces \cite{locard1930criminalinvestigation}. More broadly, this means that no interaction between two objects can occur without leaving mutual traces, even in the digital world. Systematically investigating these traces is the task of digital forensics. This computer science branch deals with the acquisition, restoration, and analysis of electronic data and includes all devices with persistent and volatile data storage. %
The digital forensic investigation is subject to high requirements to meet legal demands as digital evidence in court. Thus, an investigation must comply with the applicable government guidelines that are published from several countries and organizations \cite{BSI4n6leitfaden} from the \Gls{BSI}, \cite{NIST800_86} from the \Gls{NIST}, \cite{EU_Forensics} from the European Council, or \cite{Interpol_Forensics} from The International Criminal Police Organization – INTERPOL. %
The investigation aims to clarify the guiding questions of what happened, where, when, and how. In law enforcement and security assessment, it is also necessary to clarify who acted and whether the triggering incident can be repeated in the future. Digital forensics takes an objective role in clarifying these questions and determines both incriminating and exculpatory digital traces.
A crucial point for court standard-conform forensic analysis is the integrity and documentation of the investigation.

To guarantee this required objectivity, the \Gls{BSI} places the following requirements on the investigation process: Acceptance, Credibility, Reproducibility, Integrity, Cause and Effect, Documentation \cite{BSI4n6leitfaden}. 
Lawful documentation must also include complete proof of the digital traces' whereabouts, the chain of custody. It guarantees the authenticity of the collected data and supports the identification of the responsible party in the event of an integrity violation of the evidence trail. Therefore, integrity is of fundamental importance in digital forensics, often reached by using cryptographic hash functions.

\subsection{Digital Forensics for Smart Vehicles}
While assistant systems and smart applications in vehicles become ubiquitous, research into these vehicles' digital forensics is limited. In 2018, Le-Khac et al. presented a thorough analysis of forensic vehicle data, including a case study of a Volkswagen entertainment system \cite{smartvehicles}. While the authors make a case for using vehicle forensic data in criminal investigations, their research does not focus on smartphone applications for smart vehicles.

In 2015, Thomas Käfer \cite{kaefer} presented a security analysis of car-sharing apps and the vehicle assistant application myAudi. The analysis is in a self-published book that the author sells for 280€ at the time of this writing. From the public table of contents, Käfer analyzed the overall security of the apps and he also looked for forensically interesting data. He did not perform Subject Access Requests and he also did not specifically look into car manufacturer apps with the exception of myAudi and BMW ConnectedDrive. 

\paragraph{Investigation and Analysis of Android Auto}
In 2019, Joshua Hickman examined Android Auto -- Google's vehicle assistant application for Android -- from a digital forensics perspective \cite{android_auto}. 
Besides the date and time of the last usage of the Android Auto app, the author extracted the MAC address, the name, and the last location of the paired vehicle. Furthermore, they decoded transcribed voice recordings, dictated messages, and received instructions from the navigation from the Google Assistant app. In 2019, Sarah Edwards and Heather Mahalik were able to detect further artifacts such as log entries for phone calls and short messages sent while using Android Auto \cite{rollin_hatin}.

\paragraph{Investigation and Analysis of Apple CarPlay}

Apple's vehicle assistant application -- CarPlay -- was forensically investigated in 2019 by Binary Hick \cite{ridin_with_apple_carplay}. This analysis was extended by Sarah Edwards and Heather Mahalik in 2019 \cite{rollin_hatin}.

The researchers found that it is possible to extract the layout of the CarPlay's home screen icons, the name of the paired vehicle, the communication with the Siri voice control, short messages, stored coordinates, and an event log which shows information such as new vehicle pairings or instructions for the music player.

\section{Methodology}

To meet the requirements of digital forensics, a structured methodology must be applied. One of the best-known approaches is the \acrfull{CFSAP} model, which is geared towards the prosecution of criminal offenses \cite{cfsap}. The \acrshort{BSI} takes up this model in the digital forensics guide and defines the investigation more generally. We lean our forensic analysis of the tested smartphone applications on their guide \cite{BSI4n6leitfaden}.

Mainly, we focus our analysis on the data collection and investigation steps of digital forensics but deduct necessary information for strategic and operational preparation and the data analysis for specific cases.

For all tested vehicle app pairings, we performed the following procedure for the investigation:
\begin{enumerate}
    \item Generation of test data by running the app features on the smartphone paired with the vehicle.
    \item Data acquisition and checksum calculation of the internal memory with logged in and logged out user and after uninstalling the app.
    \item Examination and analysis of the data stock.
    \item Article 15 and 20 \Gls{sar} GDPR by the vehicle owner to the car manufacturer.
    \item Compilation of the acquired data from the IT forensic investigation and the \acrshort{GDPR} inquiry to the car manufacturer.
\end{enumerate}

\subsection{Setup}

An Apple iPhone 6s and a Xiaomi Redmi Note 4 with the latest version of iOS 13 and Android 7 at the time of the study were used. Data collection for the iPhone was carried out using the iPhone RootFS tool by James Duffy. For this purpose, a jailbreak was carried out using Checkrain 0.9.8.1 before the data collection. The memory of the Xiaomi is acquired using the alternative recovery partition \Gls{twrp} in version 3.3.1-0. The recovery partition was loaded onto the smartphone before the data collection.

The digital forensic investigation was carried out with freely available tools. The command-line tools \texttt{ls}, \texttt{cat}, \texttt{grep}, \texttt{gzip}, \texttt{file}, and \texttt{tree} on Ubuntu Linux were used to analyze the data structures. Furthermore, DB Browser for SQLite and GHex were used in addition to some individually created Python scripts for the preparation of data types.

\subsection{Generation of test data}
To carry out the digital forensic examination of different vehicle apps, we generated test data by pairing the app with the vehicle and executing the following pre-defined list of actions, if available:
\begin{enumerate}
    \item Lock and unlock the vehicle
    \item Refuel the car
    \item Drive for at least 10 minutes
    \item Start a new navigation
    \item Send a trip from the smartphone to the infotainment system
    \item Save parking position and photo
    \item Locate the vehicle
    \item Drive the car from outside by app
    \item Change the temperature of the ventilation, seat, and steering wheel heater using the app
\end{enumerate}

Following the generation of the test data, three data acquisitions of the smartphone used were carried out. The first acquisitions were carried out with the user logged into the app. After this acquisition, the logged-in user was logged out via the application, and a new acquisition was performed. The last acquisition was made after the application had been uninstalled. %
To ensure compliance with the requirements of digital forensics, we calculate the SHA-256 of all acquired images. %

The three datasets created and acquired in this way provide the basis for the subsequent investigation and analysis. The scope of our analysis is information on the key forensic questions of what happened where, when, and how and is limited to the data of the tested apps.

\subsection{Requests of Manufacturer Data}
Finally, the results of the smartphone forensics were supplemented with the data stored by the car manufacturers. The vehicle owner requested these via a corresponding \Gls{sar} according to Article 15 and Article 20 of the \acrshort{GDPR} \cite{gdpr}.

\section{Analysis}
\begin{table*}[htbp]
\small
\centering
 \begin{tabular}{ llllll }
 \toprule
 & \multicolumn{2}{c}{\textit{Android}} && \multicolumn{2}{c}{\textit{iOS}}\\
 \cmidrule{2-3}\cmidrule{5-6}
 \thead{Vehicle} & \thead{App} & \thead{Version} && \thead{App} & \thead{Version}\\
 \midrule

 Audi A4 B9             & myAudi & v3.18.0               && myAudi & v3.18.1\\
 BMW 1er F20 140i       & my BMW & v1.0.1                && my BMW & v1.0.1\\
 Ford Kuga '13          & FordPass & v3.1.0              && FordPass & v3.0.0 \\
 Mercedes C-Klasse W204 & Mercedes me Adapter & v3.11.50 && Mercedes me Adapter & v3.6.50\\
 Opel Astra K           & myOpel & v1.23.4               && myOpel & v1.23.4\\
                        & OnStar Europe & v3.28.0        && OnStar Europe & v3.28.0\\
 Seat Mii electric Plus & DriveMii App & v3.0            && DriveMii App & v3.0 \\
                        & Seat Connect & v1.1.29         && Seat Connect App & v1.1.29\\
 Tesla Model S 75D      & Tesla & v3.10.8                && Tesla  & v3.10.8\\
 Tesla Model 3          & Tesla & v3.10.9                && Tesla  & v3.10.9\\
 VW Tiguan II           & We Connect Go &v2.13.8       && We Connect Go & v2.13.6 \\

\bottomrule
 \end{tabular}
\caption{List of vehicles and tested apps. All apps were tested in late 2020.}
\label{tab:carapp}
\end{table*}

\begin{table*}[t]
\small
\settowidth\rotheadsize{\theadfont refueling}
\centering
\small
\begin{tabular}{lccccccccccccccccc}
\toprule
    & \multicolumn{8}{c}{\textit{Android}}&&
    \multicolumn{8}{c}{\textit{iOS}}\\
\cmidrule{2-9}\cmidrule{11-18}
\thead{App name}   & 
\rothead{drive log}    & \rothead{recent\\location} & \rothead{parking}      & \rothead{refueling} & \rothead{user info}    & \rothead{car info} & \rothead{logout}       & \rothead{uninstall} &&
\rothead{drive log}    & \rothead{recent\\location} & \rothead{parking}      & \rothead{refueling} & \rothead{user info}  & \rothead{car info} & \rothead{logout}       & \rothead{uninstall}\\ 
\midrule
myAudi              &\cached    & \nodata       & \nodata   & \available    & \available    & \available    & \available    & \nodata &  
                    &\cached    & \nodata       & \nodata   & \available    & \nodata       & \nodata       & \nodata       & \nodata \\ 
my BMW              & \unknown  & \available    & \unknown  & \unknown      & \unknown      & \available    & \nodata       & \nodata &
                    & \unknown  & \nodata       & \unknown  & \unknown      & \nodata       & \nodata       & \nodata       & \nodata \\ 
FordPass            & \nodata   & \unknown      & \unknown  & \unknown      & \available    & \available    & \nodata       & \nodata &
                    & \nodata   & \available    &\available & \available    & \available    & \nodata       & \nodata       & \nodata \\ 
Mercedes me Adapter &\encrypted & \encrypted    &\encrypted & \encrypted    & \cached       & \encrypted    & \unknown      & \nodata &
                    &\available & \available    &\available & \available    & \available    & \available    & \available    & \nodata \\ 
myOpel              & \unknown  & \unknown      & \unknown  & \unknown      & \available    & \available    & \available    & \nodata &
                    & \unknown  & \unknown      & \unknown  & \unknown      & \available    & \available    & \available    & \nodata \\ 
OnStar Europe       & \unknown  & \unknown      & \unknown  & \unknown      & \cached       & \available    & \available    & \nodata & 
                    & \unknown  & \unknown      & \unknown  & \unknown      & \nodata       & \nodata       & \unknown      & \nodata \\ 
DriveMii App        & \cached   & \encrypted    & \unknown  & \unknown      & \unknown      & \partialdata  & \unknown      & \nodata &
                    & \cached   & \nodata       & \unknown  & \unknown      & \unknown      & \cached       & \unknown      & \nodata \\ 
Seat Connect        & \unknown  & \nodata       & \nodata   & \unknown      & \available    & \available    & \available    & \nodata &
                    & \unknown  & \nodata       & \nodata   & \unknown      & \available    & \available    & \available    & \nodata \\
Tesla               & \cached   & \cached       & \cached   & \cached       & \cached       & \cached       & \cached       & \nodata & 
                    & \nodata   & \cached       & \cached   & \cached       & \cached       & \cached       & \cached       & \nodata \\ 
We Connect Go       &\available & \available    &\available & \available    & \nodata       & \available    & \nodata       & \nodata &
                    &\available & \available    &\available & \available    & \nodata       & \available    & \nodata       & \nodata \\ 
\bottomrule
\\

\end{tabular}
\begin{tabular}{llll}
\toprule
\available  & extensive data available      & \cached   & data cached or partially available\\
\encrypted  & encrypted database available  & \unknown  & feature not available or not tested\\
\nodata     & no data available\\
\bottomrule
\end{tabular}
\caption{Overview of the forensic analysis results.}
\label{tab:results}
\end{table*}

\sloppy

The analysis focuses on the vehicle and user data that is directly relevant for criminal investigations. The recovery of generic application data such as the app's installation date is not presented in this work but is well analyzed and described in the literature \cite{tamma2020practical}.

\cref{tab:carapp} displays the vehicles and official manufacturers' apps analyzed for this work. %
The results of the forensic analysis are summarized in \cref{tab:results}. The respective section is concluded with an explanation of the data that was disclosed by the \glspl{sar}. An overview of this data is displayed in \cref{tab:sar}. 

All iOS applications installed from the official AppStore are stored in the path \path{/private/var/mobile/Containers/Data/Application/<UUID>/}. The UUID is random and uniquely generated for every installation. Every iOS folder listed below is relative to this path. For Android applications, the default path is \path{/data/data/<package name>/}, where the package name is an application-specific ID which is the same for every installation \cite{tamma2020practical}. Every Android folder listed below is relative to this path.

\subsection{myAudi}
    The myAudi app is linked to the vehicle via a verified account and via a Blue\-tooth connection. A vehicle code issued by Audi must be entered into the vehicle to gain access to the vehicle data. With the main user set up in this way, the app can be used to read the fuel level and range, lock and unlock the vehicle, check whether the windows and hood are closed, send a planned route to the vehicle's navigation system, keep fuel and mileage logs, as well as check the vehicle's maintenance status and, if necessary, arrange service appointments.
\paragraph{iOS}
    The table \texttt{Cost\-Book\-Item} of the database \texttt{Documents/maps.db} contains the stored refueling transactions with timestamps, the refueled fuel amount, and the price paid as entered by the user. The table \texttt{DriverLogItem} stores the trips logged in the driver logbook with a start and end time and the start and destination address. Furthermore, the database table \texttt{SettingsItem} contains a timestamp indicating the last synchronization with the vehicle.
    
    After the user was logged out, the data mentioned above was not present in the database. In the dataset created after uninstalling the app, the application folder no longer exists. Thus, the previously described data of the app myAudi is deleted. However, the data is loaded from the manufacturer's backend after a re-installation and the user's login.

\paragraph{Android}
    The package of the Android version is named \texttt{en.myaudi.mobile.assistant}. In the folder \texttt{databases}, the entries of the driver's logbook are stored in the database \texttt{audiMapsDatabase.db} in the table \texttt{drivers\_log\_item}. Analogous to the iOS version, start and end time and start and destination address are stored with the corresponding mileage. In the table \texttt{cost\_book\_item}, there are entries about the performed refueling processes with a timestamp, entered price, the mileage, and the amount of fuel refueled. The \gls{JSON} file \texttt{vehicleList} in the \texttt{files} folder contains information about the vehicle, such as the model name or installed assistance systems. %
    The user's date of birth, email address, name, and user ID are stored in the \texttt{PERSISTENCE\_KEY\_USER\_ACCOUNT} file.
    We found navigation start and destination coordinates in the \path{DiskLruCache/GeoKitDecodedCoordinate/1} folder and further compressed data in the \texttt{Web\-Re\-quest\-Ma\-nagerCache} folder. This data includes historical vehicle coordinates, logs for locking and unlocking the vehicle doors, status about each separate door, information on inspections, mileage with timestamps, and the vehicle's nickname.

    In contrast to the iOS version, the determined information could be determined both in the datasets with the user account logged in and logged out. When the app was uninstalled, the application folder was completely deleted.%
\paragraph{GDPR SAR}
    The response to our \gls{sar} includes customer data such as name, date of birth, (current and previous) address, telephone number, and email address. %
    Furthermore, Audi communicates the \glspl{vin}, the model designations, the equipment, and the ownership period of all vehicles listed in the customer order history. Direct contacts with Audi AG, such as email correspondence to request the stored personal data or vehicle orders, are also communicated. The contents of the emails are not included.
    Additionally, the response includes a list of the services associated with the current vehicle, and the note that produced data from these services might be stored. %
    The actual data was not included in the manufacturer's response but can be requested separately.

\subsection{my BMW}
    The my BMW app requires the creation of a user account by entering the \gls{vin}. This sends a code to the vehicle, which must be entered into the app to verify the vehicle. Features include remote control of doors and interior ventilation, locating the vehicle, sending navigation destinations to the vehicle, an overview of various vehicle data such as fuel and mileage, and an overview of upcoming maintenance.
\paragraph{iOS}
    The app does not store relevant data on the filesystem.

\paragraph{Android}
    The Android application uses the package name \texttt{de.bmw.connected.mobile20.row}. 
    The file \texttt{.hydrated\_bloc.json} contains the year of manufacture and the \gls{vin} of the examined vehicle. Furthermore, it lists the vehicle status with the vehicle's location, timestamp, the status of the doors, and upcoming services. Additionally, the file provides information about the available features of the vehicle. 
    
    In the case of a logout or uninstallation, all data is deleted.
\paragraph{GDPR SAR}
    In addition to the customer data such as the name, date of birth, address, telephone number, and email address, the sales partners with their name and address are included in the \gls{GDPR} \gls{sar} information. Under the heading vehicle data, the history of vehicles previously registered to the customer is listed with the \gls{vin}, model designation, date of first registration, and vehicle ownership. Furthermore, correspondence with the BMW customer service is listed by date.
    
\subsection{FordPass}
    After logging into the app via a Ford account, the vehicle must be added by entering the \gls{vin}. Once the vehicle is activated in the app, a message requiring confirmation appears on the infotainment system display. Verification in this way enables the app to display mileage, tire pressure, and fuel level, list routes driven and refueling operations, and send destinations to the vehicle's navigation system. Additionally, service appointments can be made with a Ford service partner via the app.
\paragraph{iOS}
    The \texttt{Documents} folder contains several relevant SQLite databases. The \texttt{ZVEHICLE} table inside the \texttt{CoreData.sqlite} database stores the vehicle's model name, the user-assigned nickname, and the \gls{vin}.
    The \texttt{CVCoreDataModel.sqlite} database contains information about the modules installed in the vehicle, such as the head-up display or a door control module. The \gls{JSON} file in the path \path{Documents/DigitalCoPilot/dataPoints/<VIN>/snapshot} contains the fuel level with timestamps. The email address of the account used is stored in the \texttt{DTX\_8.183.1.1002.sqlite} database. %
    The file \texttt{com.ford.fordpasseu.plist} in the \path{Preferences/Library} folder contains the following data: The refueling operations with coordinates of the gas station, the last known position of the vehicle with coordinates and address, a Base64 encoded image taken to document the parking position, navigation destinations, the UserID, and metadata about the smartphone.
    
    This data is deleted in case of a user logout from the app.
    Uninstalling removes the app folder with all personal data.
\paragraph{Android}
    The Android FordPass app uses the package name \texttt{com.ford.fordpasseu}. %
    We found several empty databases with table names that appear to store forensically relevant data, such as trip destinations or vehicle position data. Since our test car did not support the newest Ford infotainment systems -- SYNC and AppLink --, we assume that these tables are only used for newer vehicle generations.
    Detailed descriptions of the vehicle are stored in the \texttt{NGSDN\_DATABASE} database. This database  contains the vehicle's exact name, the \gls{vin}, year of manufacture, and the user-assigned vehicle's nickname.
    The database \texttt{VIN\_DETAILS\_LOOKUP} stores vehicle information, such as the engine's and the transmission's model names, the warranty period, and the emission class. 
    The \texttt{com.ford.\allowbreak fordpasseu\_preferences.xml} file includes the user's account information, such as the email address, name, and \gls{vin} of the vehicle. The \texttt{com.humani\-fy.ex\-pertconnect.SHARED\_PREFS.xml} file contains an identification number and the username of the account. The \gls{xml} file \texttt{encryptedValues.xml} contains an access token, and the \texttt{pin\-Values.xml} stores the salt and the hash of the PIN of the application. These values are set during the first start-up of the app.
    
    The described information of the dataset is deleted from the memory during the logout. As with the previously mentioned apps, uninstallation of the FordPass app also deletes any personal data.
\paragraph{GDPR SAR}
    The answer to the \Gls{sar} contains only the customer name and the FordPass account's email address. The response states that the linked vehicle does not have an internet connection; therefore, no data from the vehicle is transmitted. Thus, according to Ford, there is no data from the vehicle that could be processed.

\subsection{Mercedes me Adapter}
    To start up the Mercedes me Adapter app, a Mercedes vehicle adapter must be plugged into the vehicle's \Gls{obd} interface. The adapter communicates with the smartphone via a Bluetooth connection. The account used in the app must be verified at an official Mercedes dealer by presenting the vehicle registration document and the ID card. Once this has been done, communication is possible between the smartphone app and the verified vehicle's \gls{obd} interface. The features offered include trip lists, navigation to the parked vehicle, a live overview of vehicle data such as temperatures and accelerator pedal position, a display of error messages and upcoming maintenance, and various statistics on kilometers driven and fuel consumed.
\paragraph{iOS}
    The \gls{JSON} file named \texttt{000000000000} inside the \path{Documents/1e9b5df8855a8f490f848099ef3e550c} folder records the distance of the last trip taken. %
    The file \texttt{live.json} in \texttt{Documents} contains all the information in the form of key-value pairs displayed in the app's dashboard. 
    The database \texttt{DriverLogbook.sqlite} stores the records of the driver logbook. The trips with start and destination addresses are linked to data records from the table \texttt{ZDLCOREDATRACK\-POINTS}, which stores the current vehicle location every ten seconds.
    Under \path{Library/Application Support/Live}, the database called \texttt{MBFA} contains %
    one table with the refueling processes and another with the user information. %
    
    The information described is stored on smartphone storage with both logged-in and logged-out user accounts. However, uninstalling the app deletes it from the file system.
    
\paragraph{Android}
    The Android version of the Mercedes me Adapter app stores the data under the package name \texttt{com.daimler.mbfa.android}. 
    The databases %
    \texttt{driverlogbookDatabase.db} and \texttt{mbfa.db} are encrypted. Since databases with the same name could be examined in the iOS version, it can be assumed that similar information is stored in them. However, this could not be proven as we were unable to decrypt the database files.
    In the folder \texttt{app\_webview}, there is a database called \texttt{Web Data}. The table \texttt{autofill} contains the address of the account holder and the timestamps \texttt{date\_created} and \texttt{date\_last\_used}. The \texttt{cache} folder contains \texttt{com.google.android.gms.maps.vol\-ley} and \texttt{vol\-ley} files, each of which contains \glspl{url} -- these show map sections from Google Maps displayed in the application during data collection. The \texttt{resources} folder contains a photo taken by the user via the app to record the vehicle's parking position. In the \texttt{shared\_prefs} folder, there are several \gls{xml} files. The \texttt{com.daimler.mbfa.android\_preferences.xml} file contains, among other things, the \gls{vin}, the \gls{obd} adapter's identification number, and the remaining kilometers till the next recommended service appointment.%
    
    Analogous to the iOS version of the app, all information remains available after the user logs out. The entire folder structure of the app is deleted when the app is uninstalled.
\paragraph{GDPR SAR}
    The information provided by Mercedes-Benz AG in response to the \Gls{sar} included the customer's name, date of birth, email address, and address. Furthermore, it can be seen from the information at what time the user conditions were agreed to by email, SMS, or letter. No information about the vehicle is provided. %

\subsection{myOpel}
    The myOpel app uses a myOpel account, which can be linked to the vehicle by entering the \gls{vin}. Verifying the vehicle in the myOpel account is done via an Opel dealer by presenting the vehicle registration document. With the tested Opel Astra K, only the mileage could be viewed via the app, and a service appointment could be made. Further features were not available.
\paragraph{iOS}
    In the \texttt{Documents} folder, log files under \path{LogDirectory/com.psa.myopel} can be used to determine user information such as the email address and the \gls{vin}. The database \texttt{UserProfileModel.sqlite}, also stored under \texttt{Documents}, contains user information and information about the connected vehicle in addition to the dealer specified during setup. The data structure of the \texttt{BTAModel.sqlite} database suggests that it records traveled routes with timestamps and location information. Due to the limited range of functions of the available vehicle, there are no corresponding entries in the investigated dataset. However, the database \texttt{BOUserMyMarqueModel.sqlite} contains entries for the warranty, email address, and the \gls{vin}. In the \texttt{Library} directory, under \texttt{Preferences}, there is a \gls{plist} file with user information and the vehicle's \gls{vin}. It also contains coordinates of the smartphone with a corresponding timestamp.
    
    The information shown is available in the dataset with the user logged in and logged out. During the uninstallation, the complete folder with all listed data is deleted.
\paragraph{Android}
    \sloppy
    The Android version of the app myOpel uses \texttt{com.psa.mym.myopel} as the package name. %
    The \texttt{cache/logs} folder contains text files that include the vehicle's \gls{vin}. 
    The \texttt{cache/WebView/Default/HTTP Cache} stores cached communication between the app and the manufacturer's backend. From our dataset, we were able to reconstruct the loading of the vehicle manual from the server.
    The \texttt{database} folder contains the \texttt{BOUserMymarque.db} database known from the iOS analysis, with a slightly different name. Analogous to the iOS version, the email address used, the \gls{vin}, and the guarantee status can be viewed. The databases \texttt{CarProtocolStrategy.db}, \texttt{LocalisationSmartphone.db}, \texttt{SmartAppsV1.db}, and \texttt{Smart\-Apps\-V2.db} each contain tables that are presumably designated for storing coordinates. However, these tables are empty in the datasets examined. The database \texttt{UserProfile.db} stores the same information as the database \texttt{UserProfileModel.sqlite} of the iOS version. This includes user and vehicle information as well as the registered service partner. In the \texttt{shared\_prefs} folder, an \gls{xml} file called \texttt{com.psa.mym.myopel\_preferences.xml} contains the vehicle's \gls{vin}, the user account's email address, and the settings made for the application.
    
    By comparing the datasets, it could be determined that the information presented is not deleted by logging off the user. This is only the case by uninstalling the app.
\paragraph{GDPR SAR}
    Opel provides the name of the customer as well as the address, telephone number, and email address for the \gls{GDPR} inquiry. For the linked vehicle, the registration number, \gls{vin}, and model designation are listed.

\subsection{OnStar Europe}
    An account must be created to use OnStar Europe by requesting an OnStar account number via the vehicle's OnStar button.
    After activation, vehicle data can be retrieved via the app or the OnStar web app. The information displayed includes the current mileage, tire pressure, fuel level, time until the next oil change, and the vehicle's current location. Furthermore, the vehicle can be opened and closed remotely, a WiFi hotspot can be activated, and the lights and horn can be operated. When the data collection was carried out, only the mileage could be displayed via the app, and the parking position could be set and saved with a picture. Further information that was displayed via the web application was not available via the app. The app also indicated that it would no longer be available after 31.12.2020 and that the user should switch to the myOpel app.
\paragraph{iOS}
    The app does not store relevant data on the filesystem.
    
\paragraph{Android}
    OnStar Europe is listed in the Android file structure with the package name \texttt{com.gme.opel.owner.android}. %
    The \path{cache/GeminiCache} folder contains downloaded gzip archives with \gls{JSON} files. The archive \texttt{16d8b336686532339c0e35c784c68215.1} contains information about the vehicle, such as the \gls{vin} and the model designation. The SQLite database \texttt{mylink} in \texttt{databases} contains 
    information about the vehicle, such as the model name in the table \texttt{vehicles} and the tire pressure and mileage in the table \texttt{vehicle\_diagnostics}. Other tables, presumably documenting parking positions and routes were empty in the datasets examined. %
    
    Both with the user logged on and logged off, the information presented in this section could be viewed. During uninstallation, the complete application folder is deleted.
\paragraph{GDPR SAR}
    In response to the \gls{sar} to OnStar Europe, the customer data, such as name, address, telephone number, email address, and date of account creation are provided. Furthermore, the \gls{vin} and the mobile phone identifiers such as \gls{imsi}, \gls{imei}, \gls{iccid}, and \gls{msisdin} are provided for the linked vehicle. Also, it is shown at which time the customer used which product packages. OnStar product packages include extended service offers or monthly vehicle checks based on transmitted data. A log records the time at which warnings were sent. In our case, these include the messages \texttt{oil} and \texttt{low tires}.
    
    Additionally, phone calls with OnStar advisors via the in-vehicle button are logged, including the vehicle’s coordinates and the start and end.
    This conversation history goes back to a time when the applicant was not yet the owner of the vehicle. Emails sent and marketing campaigns carried out are listed without content.%

\subsection{DriveMii App}
    Seat's DriveMii app is designed to complement the vehicle's infotainment system with offline navigation, an Ecotrainer that displays the energy consumed and recuperated, a media player that can play music from the smartphone, and information about the current trip. For this purpose, the smartphone can be connected to the vehicle via Bluetooth. %
    Account registration is not necessary.
\paragraph{iOS}
    The \gls{vin} of the paired vehicle can be recovered from the file \texttt{com.seat.connected\-car.drivemii.plist} in the folder \texttt{Library/Preferences}. The table \texttt{ZRECUPERATIONHIS\-TO\-RY} of the SQLite database \texttt{ElectricalService.sql} in the folder \texttt{Documents} stores records about the recuperation created every minute.
    
    We found several, apparently encrypted, databases in the folder \texttt{Application Support/com.seat.connecte\-car.drivemii/home}. From the files' names, it can be assumed that relevant data such as location, itinerary, and track data is stored in the following databases: \path{fav/locations.sqlite}, \path{fav/markers.sqlite}, \path{itn/itineraries.sqlite}, and \path{tracks/tracks.sqlite}.
    The file \texttt{DE\_AT\_CH\_MapSettings\_.tlv} in the same folder contains the user-entered navigation destinations, such as the town or street name.  They are stored in plain text but without any timestamps or further information.
    
    Since the app does not offer a login option, only the datasets with the app installed and uninstalled could be compared. When the app is uninstalled, the information described is no longer available.
\paragraph{Android}
    The Android version of the DriveMii app stores the data under the package name \texttt{com.seat.connectedcar.drivemii}. The folder \texttt{files} contains the encrypted databases in the same way as the directory structure of the iOS app under \path{data/files}. %
    Comparable to the \texttt{DE\_AT\_CH\_MapSettings\_.tlv} of the iOS app, the \texttt{DE\_AT\_CH-1026 66\_MapSettings\_.tlv} file on Android contains the navigation inputs. The \texttt{App4EntryPrefs.xml} file inside the \texttt{shared\_prefs} folder lists the application settings, such as the app's icon arrangement, and stores the \gls{vin} of the connected vehicle.
    
    Since the Android app does not require a user account, no data collection could be carried out after logout. Uninstalling the app deletes the previously displayed information from the storage.
    
\paragraph{GDPR SAR}
    See the next section for the \gls{sar} sent to Seat.
    
\subsection{Seat Connect}
    The Seat Connect app uses a Seat account, which must be linked to the vehicle via the \gls{vin}. 
    When the user is logged in, the app displays the remaining range till recharge, whether the doors and windows are locked, whether the lights are switched off, the average energy consumption, and the time until the next scheduled service.
    Furthermore, the batteries' charging can be planned, the vehicle's current position can be determined, and the air conditioning can be controlled remotely.
\paragraph{iOS}
    Inside the application's directory structure, the folder \texttt{Library/Pre\-ferences} contains the \gls{plist} file \texttt{com.seat.connectedcar.mod3connectapp.plist}. The key-value pairs contained therein include, among other information, a timestamp of the last user login, the vehicle's \gls{vin}, meta-information about the smartphone used, and user information such as phone number, date of birth, and email address.
    
    The information shown is also available in the dataset when the user account is logged out. Uninstalling the app deletes the entire application folder.
\paragraph{Android}
    The Android version of the app uses the package name \texttt{com.seat.connectedcar.mod2connectapp}. %
    The SQLite database under \texttt{Default/Web Data} stores a table \texttt{autofill}, where the user account's email address is stored. In the folder \texttt{databases}, the \texttt{ModAppDatabase.db} contains the table \texttt{Persis\-tentUser} with the user account's email address, the \gls{vin}, and the assigned vehicle nickname connected to the app. The table \texttt{PersistentVe\-hicleMetadata} contains the vehicle's exact name and its \gls{vin}.
    
    The listed databases remain with the records after the user logs out. %
    Uninstalling the app removes the complete application folder.
\paragraph{GDPR SAR}
    In response to the \gls{sar}, Seat provided the customer's name, date of birth, address, nickname, phone number, and email address. Furthermore, any vehicle access is logged, which allows a reconstruction of the individual vehicle usages.

\subsection{Tesla}
    After downloading the Tesla app, the user is prompted to log in via their Tesla account. Mobile access must be allowed in the vehicle in order to access the vehicle from an app. Once this is done, the range and temperatures of the vehicle can be read in the app. In addition, the lights and the vehicle itself can be switched on and off, doors can be locked and unlocked, the hood and trunk can be opened, the interior temperature and seat heating can be regulated, and the vehicle can be made to sound its horn. Also, the vehicle can be driven forward and backward from outside via the app. To do this, the user must be within proximity of the vehicle.
\paragraph{iOS}
    We analyzed the app with two different cars and two app versions. The data of version 3.10.8 was generated with a Tesla Model S 75D, whereby version 3.10.9 was tested with a Tesla Model 3. The outcome of both analyses is the same.
    
    The table \texttt{cfurl\_curl\_re\-ceiv\-er\_data} of the database \texttt{com.teslamotors.TeslaApp}, stored in the file \texttt{Library/Caches/Cache.db}, contains the vehicle's static data, such as the vehicle name and the \gls{vin}.
    Dynamic data, namely the coordinates of the last location, the vehicle status including the interior temperature, and the state of charge with timestamps, is stored in the \gls{JSON} file \texttt{Library/Caches/fsCachedData}.
    
    Logging out does not delete the data and is still available for recovery. When the app was uninstalled, the complete dataset of the examined folders is deleted.
    
\paragraph{Android}
    The Android version of the Tesla app uses the string \texttt{com.tesla\-motors.tesla} as the package name. %
    The folder \texttt{app\_webview} contains an SQLite database under \texttt{Web Data}, in which the user's email address is located in the table \texttt{autofill}. 
    In the folder \texttt{http-cache}, data packages in \gls{JSON} format are found that contain the \gls{vin}, user ID, and the vehicle status. The most important information of the status includes the vehicle's location, the speed, and the gear enriched with timestamps. Other \gls{JSON} files contain the user's first and last name, the account name, which corresponds to the email address, various configurations and equipment of the vehicle, and the vehicle's name.%
    
    The explained information from the examined datasets could be determined both with the Tesla account logged in and logged out. The uninstalling process deleted the folder \texttt{com.teslamotors.tesla} with the data stored in it. A difference between the app versions 3.10.8 and 3.10.9 could not be determined.
\paragraph{GDPR SAR}
    The data provided by Tesla in response to the \Gls{sar} shows that, in addition to the vehicle owner's information, data from the vehicle is transmitted to Tesla. The vehicle owner's information includes name, address, phone number, email address, orders for vehicles, and services performed. In addition to the information on the model name, the \gls{vin}, warranty, and the vehicle's number plate, tables with up to 229 columns for six days were included in the information provided. These are labeled \texttt{Accelerator Pedal Position (\%)}, \texttt{Estimated Brake Pedal Position}, \texttt{Autosteer Driver Hands On Detection}, and \texttt{Primary Steering Angle Sensor (degrees) (Positive indicates right turn)}, thus describing the vehicle's conditions in detail. The tables' records show, based on the timestamp, that they are collected up to ten times per second. An attached document states that the regularly collected data may be transmitted to Tesla's servers to ensure the vehicle's continuous performance and predictive maintenance. The reasons why the six days, in particular, were chosen and why different numbers of data fields were transmitted to Tesla are not explained.  

\subsection{We Connect Go}
    To use the We Connect Go app, a corresponding VW DataPlug must be plugged into the vehicle's \gls{obd} interface, similar to the Mercedes me Adapter. Provided that registration has taken place via a VW account, a connection to the VW DataPlug can be established via the smartphone's Bluetooth interface using the code printed on the DataPlug. After successful pairing, the app displays the fuel level and range, current mileage, and other vehicle data. The app offers the possibility to keep fuel and trip logs and to arrange service appointments. Furthermore, the app attempts to motivate the driver to adopt a more economical driving style through challenges and offers additional statistics and driving style analyses.
\paragraph{iOS}
    For the iOS app, we recovered a parking position image, taken with the app, inside the folder \texttt{Documents/.avacar\_SUPPORT/\_EXTERNAL\_DATA}. Exact names and identification numbers for the VW DataPlug can be found in the database \texttt{VW\_DataPlug\_2\_1\_ClientURLTranslation\_5\_1.sqlite3}.
    
    We identified the file \texttt{avacar.db} as the main SQLite database. This database stores extensive data that is relevant for forensic investigations in several tables. This data includes detailed static vehicle information such as the \gls{vin}, model code, engine, and transmission designation. Furthermore, statistical data such as average fuel consumption and the fuel level with timestamps is processed by the database. The stored dynamic trip data comprises refueling processes with coordinates and timestamps, traveled distances, start and destination addresses with coordinates and timestamps, and acceleration and deceleration values with corresponding velocities. Finally, the parking positions are listed with coordinates and timestamps.

    The described main database \texttt{avacar.db} is deleted in the case of a user logout. The remaining data, namely the parking position image and the DataPlug information, is still recoverable.
    The complete app folder is deleted from the file system after uninstalling the app.
    
\paragraph{Android}
    The We Connect Go app uses the name \texttt{en.volkswagen.vwconnect}. %
    The directory \texttt{database} contains the main database \texttt{avacar.db} known from the analysis of the iOS version and has the same structure. Thus, analogous to the analysis of the iOS version, detailed vehicle and trip data is collected.
    
    A logout deletes all data from the database \texttt{avacar.db}. After uninstalling, the examined folder \texttt{en.volkswagen.vw\-connect} is completely removed.
    
\paragraph{GDPR SAR}
   In response to the \gls{sar}, Volkswagen AG discloses the customer data with name, date of birth, address, email address, telephone numbers stored, and the vehicle's \gls{vin} and registration number. Further processed personal data  is divided into different categories. According to this, Volkswagen stores, among other things, Car2X, accident, contract, and position data. %
\fussy

\begin{table}[t]
\small
\settowidth\rotheadsize{\theadfont \textbf{Correspondence}}
\centering
\small
\begin{tabular}{lccccccc}
\toprule
\thead{Manufacturer}   & 
\rothead{Customer Data}    &
\rothead{Vehicle Data}&
\rothead{Infotainment\\Usage} &
\rothead{Correspondence}    &
\rothead{Order History}      &
\rothead{Position Data} &
\rothead{Additional Data}       \\
\midrule
Audi                & \available    & \available    & \available    & \metadata & \partialdata  & \nodata       & \nodata       \\ 
BMW                 & \available    & \available    & \nodata       & \metadata & \nodata       & \nodata       & \nodata       \\
Ford                & \available    & \nodata       & \nodata       & \nodata   & \nodata       & \nodata       & \nodata       \\
Mercedes            & \available    & \nodata       & \nodata       & \nodata   & \nodata       & \nodata       & \nodata       \\
Opel                & \available    & \available    & \nodata       & \nodata   & \nodata       & \nodata       & \nodata       \\
OnStar              & \available    & \additional   & \available    & \metadata & \available    & \partialdata  & \additional    \\
Seat                & \available    & \available    & \available    & \nodata   & \nodata       & \nodata       & \nodata       \\
Tesla               & \available    & \available    & \nodata       & \nodata   & \available    & \available    & \additional   \\
Volkswagen          & \available    & \available    & \nodata       & \nodata   & \available    & \available    & \additional   \\
\bottomrule
\\

\end{tabular}
\begin{tabular}{llll}
\toprule
\available  & data available        & \partialdata  & partial data available\\
\metadata   & metadata available    & \nodata       & no data available\\
\additional & \multicolumn{3}{l}{extensive data available (see text)}\\
\bottomrule
\end{tabular}
\caption{Overview of data included in GDPR SAR responses.}
\label{tab:sar}
\end{table}

\section{Evaluation and Discussion}

Our investigation shows that the apps handle data, such as locations logged by keeping logbooks and saving parking positions, in diverse ways, as displayed in \cref{tab:results}. While only the Mercedes Me adapter iOS application left extensive data of all categories on the phone storage, all tested apps left forensic data traces of some variation.%

Almost all of these data traces could aid in criminal investigations, e.g., to prove that a suspect interacted with the vehicle during the time of the suspected wrongdoing, essentially proving or disproving an alibi. The data becomes even more relevant when location data is logged -- for navigation or parking -- as was the case, for example, in the Mercedes Me adapter app.

The scope of the data stored on the smartphone strongly depends on the equipment of the vehicle. In the case of the Mercedes and Volkswagen vehicles examined, additional hardware was required for data transmission. Vehicles from Mercedes or Volkswagen that do not require this additional hardware were not available. However, the vehicles examined in which communication between the vehicle's \acrshort{obd} interface and the smartphone occurs via Bluetooth provided the most extensive data through the digital forensic analysis. This suggests that in these cases, the smartphone acts as a repository for the data collected. In the cases investigated where the vehicle's \acrshort{gsm} interfaces were used, only data retrieved in the app was cached on the smartphone. In this case, the underlying data stock is not completely on the smartphone but the manufacturer's server or in the vehicle itself.

Interestingly, the analysis of the apps on both iOS and Android shows that the data storage differs significantly between both versions. For example, persistently stored data from the Mercedes Me adapter app is kept in databases in both the iOS and Android versions. However, in contrast to the Android version, the app's iOS version did not encrypt any data. Thus, the data collected on the iOS version could be used for further investigation.

The manufacturers' responses to the vehicle owners' \glspl{sar} also differed significantly between manufacturers concerning the data provided, as displayed in \cref{tab:sar}. Except for Mercedes -- which only provided the customer data on file -- the customer data and the vehicles linked to the customer were mentioned in all of the responses.
The included vehicle information ranged from naming the model and the associated \acrshort{vin} to a detailed list of the vehicle's equipment. Audi and Volkswagen state in their information what data they collect on the vehicle but did not share it in response to the \acrshort{sar}. The data mentioned range from location data and data from control units to the vehicle's status and the user's Twitter account. An attached form can be filled out at Volkswagen, to request a copy of the data. Tesla provides the data recorded by the vehicle and transmitted to Tesla in an extensive spreadsheet to the applicant. OnStar Europe communicated when and from which location communication to the OnStar service was established. The data even included a period when the applicant was not yet the vehicle owner -- which is a serious privacy violation. The information provided by OnStar Europe also states that emergency calls and warnings, for example of low tire pressures, are logged by the service provider in conjunction with the vehicle's location data.

Tesla's extensive data shows the potential benefit law enforcement agencies could reap from requesting customer data from the car manufacturer. Since Tesla includes data on 229 metrics with an accuracy of up to 10 times per second, almost any actions performed with the vehicle can be reconstructed. In combination with other evidence, this data could aid in reconstructing criminal cases, including checking alibis and even convicting or exonerating the driver of a crime involving the vehicle. While only Tesla provided such extensive data, the data we received in response from other manufacturers might again be limited by the tested vehicles' features and equipment. We expect that more modern vehicles will provide even more data to the manufacturer and the assistant application, providing additional useful information for criminal investigations.

\section{Conclusion}
This work shows that the extensive data generated by modern vehicles is transmitted to car manufacturers and transferred and stored on drivers' smartphones through vehicle assistant apps.

The digital forensic examination of the apps shows that this data is a useful addition to vehicle forensics. Thus, using digital forensic methodology, it was possible to analyze how vehicle data stored on the smartphone can be retrieved in order to contribute to clarifying the key digital forensic questions. 

The information provided by car manufacturers in response to the \glspl{sar} shows that often only data on the vehicle's owner, designation, and the \acrshort{vin} are present. Extensive data, such as that stored in Mercedes and Volkswagen's apps, was only provided by Tesla. Thus, it can be concluded that this personal data is not within the direct access of the car manufacturers and is stored in the vehicles' storage. The vehicle data provided by Tesla shows through the volume and diversity of the transmitted data the possibilities of data collection available in modern vehicles. This vehicle data collected on a millisecond basis and partially transmitted to Tesla offers considerable support in evaluating and reconstructing incidents. 

Our research results provide a positive result to our research question about the usefulness of forensically acquired data from vehicle assistant apps and data requested from manufacturers in investigating criminal offenses. Such extensive data as acquired from Tesla and the forensic analysis of various apps could be a crucial aid in solving criminal offenses. While not all tested apps and contacted manufacturers provide such extensive data, the trend shows that the amount of data collected by future vehicle assistant systems will increase, providing even more data to investigators.

In summary, this work provides insights to be considered for the strategic preparations of future digital forensic investigations in the field of vehicle forensics. In particular, we recommend that law enforcement considers the data generated by vehicle assistant applications to aid in investigations.

\subsection*{Future Work}

During the forensic analysis, we found several factors impeding the data collection. For one, the encryption of databases in Android applications -- i.e., the Mercedes and DriveMii app -- prevents access to the collected information. Further forensic research should look into the encryption's strength and key management as there may be ways to decrypt this data.

Another factor was the caching of data. While we were able to reconstruct cached data in applications after usage, we did not analyze the actual caching behavior. Specifically, we did not test how long data stays in the cache. It would be interesting to learn how long data remains available for forensic analysis.

\section*{Acknowledgments}
The authors would like to thank Christof Wetter and others for providing test vehicles for our analysis setup.

Fabian Ising and Christoph Saatjohann were supported by the research project ``MITSicherheit.NRW'' funded by the European Regional Development Fund North Rhine-Westphalia (EFRE.NRW). Fabian Ising was also supported by a graduate scholarship of Münster University of Applied Sciences.

\bibliographystyle{ACM-Reference-Format}
\bibliography{references}
\end{document}